\documentclass[%
reprint,
superscriptaddress,
showpacs,preprintnumbers,
showkeys,
 amsmath,amssymb,
 aps,
]{revtex4-1}

\usepackage{graphicx}
\usepackage{dcolumn}
\usepackage{bm}

\def\be{\begin{equation}}
\def\ee{\end{equation}}
\def\bea{\begin{eqnarray}}
\def\eea{\end{eqnarray}}
\newcommand{\ket}[1]{\mbox{$|#1\rangle$}}
\newcommand{\bra}[1]{\mbox{$\langle#1|$}}
\newcommand{\avg}[1]{\mbox{$\langle#1\rangle$}}

\begin{document}

\title { Electrical Control of  Strong Spin-Phonon Coupling in a Carbon Nanotube }

\author{Fang-Yu Hong}
\affiliation{Department of Physics, Center for Optoelectronics Materials and Devices, Zhejiang Sci-Tech University,  Hangzhou, Zhejiang 310018, China}

\author{Jing-Li Fu}
\affiliation{Department of Physics, Center for Optoelectronics Materials and Devices, Zhejiang Sci-Tech University,  Hangzhou, Zhejiang 310018, China}

\author{Yan   Wu}
\affiliation{Department of Physics, Center for Optoelectronics Materials and Devices, Zhejiang Sci-Tech University,  Hangzhou, Zhejiang 310018, China}

\author{Zhi-Yan Zhu}
\affiliation{Department of Physics, Center for Optoelectronics Materials and Devices, Zhejiang Sci-Tech University,  Hangzhou, Zhejiang 310018, China}
\date{\today}
\begin{abstract}
We describe an approach to electrically control the strong interaction between a single electron spin and the vibrational motion of a suspended carbon nanotube resonator.  The strength of the deflection-induced spin-phonon coupling is  dependent on the wavefunction of the electron confined in a lateral carbon nanotube quantum dot. An electrical field along the nanotube shifts the effective center of the quantum dot, leading to the corresponding modification of the spin-phonon strength.  Numerical simulations with experimentally reachable parameters show that  high fidelity quantum state transfer between mechanical and spin qubits driven by electrical pulses is feasible. Our results form the basis for the fully electrical control of the coherent interconvertion between light and  spin qubits and for manufacturing electrically driven quantum information processing systems.
\end{abstract}


\maketitle
\section{INTRODUCTION}

 Nanomechanical  devices have attracted broad interest as they hold promise to find potential applications in  fundamental tests of  quantum behavior of macroscopic objects, ultrasensitive  detection, and  quantum information processing \cite{mphz}. The exceptional mechanical properties of carbon nanotubes (CNTs), their large Young modulus, and low masses on the order of the attogram, make  them to be the most conspicuous among all kind of available  nanomechanical systems, because these properties  result in high resonance frequency and large zero-point motion, which make it possible to greatly enhance the quantum operation speed and to relieve the  demanding experimental conditions for ground-state cooling and state readout.

 Moreover, carbon-based systems have many fascinating features for quantum information processing.
In a variety of systems like GaAs, the hyperfine interplay between electron and nuclear spins contributes the leading source of electron spin decoherence, which heavily hinders the qubit performance of the single electron spins in a quantum dot defined in these systems. In contrast, we can grow carbon-based systems  with pure $^{\text 12}$C which is free of net nuclear spin, to completely  remove the hyperfine-interaction induced decoherence. Moreover, compared with phonon continuum in bulk materials which imposes main restrictions on the improvement of the electron spin coherence time, the phonon spectrum of a suspended CNT is discretized and can be designed to have a a remarkably low density of states, enabling very long spin lifetimes when the spin energy splitting is off-resonance with the phonon modes. At the same time we can  obtain strong spin-phonon coupling  for  quantum information processing such as qubit rotation, quantum state transfer, or generating entangled states,  by prompting the spin splitting resonant with a high-Q discrete phonon mode.

Quantum control of the state of an electron spin is generally accomplished by a resonant oscillating magnetic field. However, it has been proven to be challenging to generate strong oscillating magnetic field in a semiconductor device by specially designed  microwave cavities \cite{bsps} or microfabricated striplines \cite{fkcb}. In contrast, it is far easier to produce electric field by simply exciting a local gate electrode. Furthermore, electric field provides much higher spatial selectivity, which is essential for local individual spin manipulation. Electrical control of a spin-based system for quantum information processing \cite{knfk, gsyk} is thus highly desirable. Recently literature \cite{kfcm} theoretically  shows that electric spin control and coupling is  viable with bends in CNTs. The deflection in a CNT  can also lead to  strong coupling between an electron spin in a quantum dot and the vibrational motion of the CNT \cite{apps,cocs}. However the suggested spin-phonon coupling is difficult to modulate when the setup has been shaped up.

 Here we propose a scheme for electrical control of the spin-phonon coupling in  CNTs.
   The deflection due to the vibration motion of a suspended CNT couples the CNT vibration motion  to an electron spin confined in a lateral CNT  quantum dot (QD).    The spin-phonon coupling is  dependent on the electron density distribution in the QD. An electrical field along the nanotube shifts the effective center of the QD, thus  modifies the density distribution,  leading to the corresponding electrical control of the spin-phonon strength. Note that the spin-phonon coupling can be easily turn off through the electric field, eliminating the phonon mechanism of decoherence. Numerical simulations show that the high fidelity quantum state transfer between mechanical and spin qubits driven by electrical pulses is within the reach of current technique. Recent experimental advance in fabrication of lateral CNT QD provides  the basis for the physical implementation of our proposal \cite{jsjg}.   Our results provide the basis for the electrical control of light-spin quantum interface and for manufacturing electrically driven quantum information processing systems.

\begin{figure}[t]
\includegraphics[width=8cm]{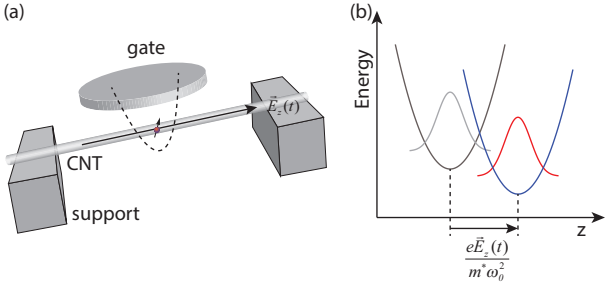}
\caption{\label{fig1}(color online) (a) Schematic of a lateral  quantum dot confining an electron spin on a suspended carbon nanotube. (b) A parallel-to-CNT electric field  displaces the effective center of the parabolic potential along the electric field direction and changes the potential depth, $\omega_0$ is the oscillator frequencies of the parabolic quantum dot, and $m^*$ is the effective electron mass. see text for details.}
\end{figure}

\section{Theoretical model and numerical evaluation}
In the setup shown in Fig. \ref{fig1} the suspended CNT confines an electron with a lateral quantum dot. Here we consider most experimentally relevant regime $\Delta_g\gg E_L\gg E_B, E_{\text {SO}}, E_{KK'}$, where
$2\Delta_g$ is the energy gap between the valence band and the conduction band, $E_L$ is the energy level difference
arising from the longitudinal confining potential of the quantum dot, and $E_B$, $E_{\text {SO}}$, and $E_{KK'}$  are
the energy changes caused by the external magnetic field, the spin-orbit coupling, and the intervalley interplay,  respectively \cite{appss}. If a  magnetic field of magnitude around
\be\label{eq1}
B^\ast=\frac{\Delta_{\text{so}}}{2\mu_B}\sqrt{1-\frac{4\Delta^2_{KK'}}{\Delta_{\text{so}} (\frac{\mu_{\text{orb}}}{\mu_B^2}-1)}},
\ee
is applied along the longitudinal direction of the CNT, the system shown in Fig. \ref{fig1} can be treated as a spin qubit coupled to a nanomechanical oscillator mode with the following Hamiltonian \cite{apps}
\be\label{eq2}
H=\frac{1}{2}\omega_s\sigma_3+g(a+a^\dagger)\sigma_1+\omega_r a^\dagger a,
\ee
where $\mu_{orb}$ and $\mu_B$ denote the electronic orbit and spin magnetic moments, $\Delta_{\text{so}}$ and $\Delta_{KK'}$ are the spin-orbit and intervalley couplings,  $\omega_s$ is the tunable spin qubit splitting,  and $\omega_r$ is the frequency of the fundamental bending mode of the CNT described by  the annihilation operator $a$. $\sigma_{1,3}$ are Pauli matrices operating on the two-level spin spaces.
Here the spin-phonon coupling
 \be
 g=\Delta_{\text{so}}\avg{f'}u_0/2\sqrt{2}
 \ee
 is independent of the magnetic field $B$ along the CNT, $\avg{f'}\equiv \int dz\frac{df(z)}{dz}n(z) $ is the average  of the derivative of the waveform $f(z)$ of the phonon mode against the electron density profile $n(z)$ in the QD, and $u_0$ is the amplitude of the zero-point fluctuations of the phonon mode.

\begin{figure}[t]
\includegraphics[width=8cm]{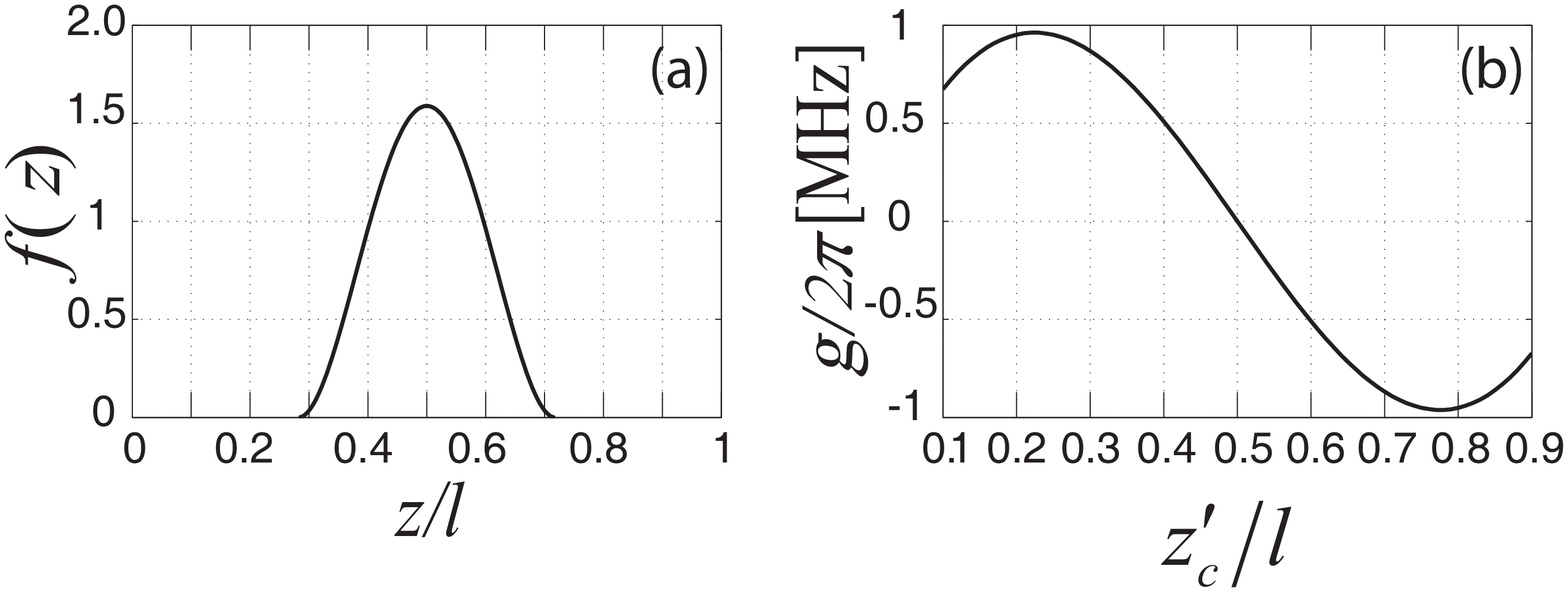}
\caption{\label{fig2} (a) The dimensionless waveform $f(z)$ of the phonon mode under consideration with normalization $\int_0^l f^2(z)dz=l$. (b) The spin-phonon coupling $g$ versus the effective center of the lateral CNT QD $z'_c$.  see text for details.}
\end{figure}
The first flexural eigenmode of the  CNT clamped at both sides $z=0$ and $z=l$ has the form \cite{mphz}
\be \label{eq3}
f(z)=\cosh kz-\cos kz+\frac{\cos\beta_0-\cosh\beta_0}{\sin\beta_0-\sinh\beta_0}(\sin kz-\sinh kz),
\ee
where $\beta_0=4.730$ and $k=\beta_0/l$.  $f(z)$ is normalized to $\int_0^l f^2(x)dx=l$. The confinement  potential for the lateral QD embedded in a CNT can be modelled as a  parabolic potential \cite{srmm}
\be \label{eq4}
\frac{1}{2}m^\ast\omega_0(z-z_c)^2,
\ee
where $m^\ast$ is the effective mass of the electron and $\omega_0$ is the characteristic harmonic oscillator frequency, $z_c$ is the center of the parabolic well. Assuming that the electron in the lateral CNT QD is in the ground state, the electron density profile $n(z)$ has the form \cite{asmr}
\be\label{eq5}
n(z)=\frac{\alpha}{\sqrt{\pi}}\exp[-\alpha^2(z-z_c)^2],
\ee
where $\alpha=\sqrt{m^\ast\omega_0}$.

Because the spin-phonon coupling $g$ is dependent on the electron density profile $n(z)$, we can modulate the spin-phonon coupling strength by  altering the electron distribution in the QD. For this reason an electric field is applied along the longitudinal direction of the CNT (Fig. \ref{fig1} a). This electric field does not modify the energy level spacing of the quantum dot, but does shift the effective center of the quantum dot to $z'_c=z_c+eE_z/(m^\ast\omega_0^2)$ \cite{jdwa}.

As an example we consider the case where the QD is located at the midpoint of the CNT, $z_c=l/2$. With realistic parameters at hand \cite{gsah,fksi}, $l=400$ nm, $u_0=2.5$ pm, $\Delta_{\text{so}}=370\mu$eV, $\alpha=40/l$, we find the relation between  the spin-phonon coupling $g$ and the effective center of the QD shown in Fig.\ref{fig2}b. The maximum magnitude of $g$ is about 0.961 MHz.

  Carbon nanotube resonators with high resonance frequencies of several hundred megahertz and a high Q exceeding $10^5$ have been reported in \cite{gsah,ahgs}. For a resonator of frequency $\omega_r/2\pi=500$MHz and mechanical quality factor $Q_r=1\times10^5$ we have a resonator damping rate $\gamma_r/2\pi=\omega_r/ Q_r=5\times 10^4$ Hz$\ll g$. Even in bulk materials such as  a one-electron GaAs QD literatures \cite{sakm, rhlk}  report spin dephasing rate $\gamma_s$ of several Hz, at the same time near-zero density of state of other phonon modes of a CNT  at $\omega_s$ can further lower the spin dephasing rate, resulting in a reasonable assumption $\gamma_s\ll g$. Thus the spin-mechanical coupling strength $g$ far exceeds the damping rate $\gamma_r,\gamma_s$, bringing the hybrid system into the so-called ``strong coupling" regime.

In the following discussions  the resonance condition, $\omega_q=\omega_r$ is assumed. Applying the rotating wave approximation (RWA)we can rewrite the Hamiltonian Eq.(\ref{eq2}) into the  Jaynes-Cummings form  in the interaction picture
\be \label{eq6}
H_{\text {in}}=ga\sigma_++ga^\dagger\sigma_-,
\ee
 where Pauli matrices $\sigma_+=\ket{1}\bra{0}$ and $\sigma_-=\ket{0}\bra{1}$ with spin qubit states $\ket{0}$ and $\ket{1}$. Hamiltonian \eqref{eq6} allows a coherent quantum state transfer between the spin and  nanomechanical resonator modes.

  \begin{figure}[t]
\includegraphics[width=8cm]{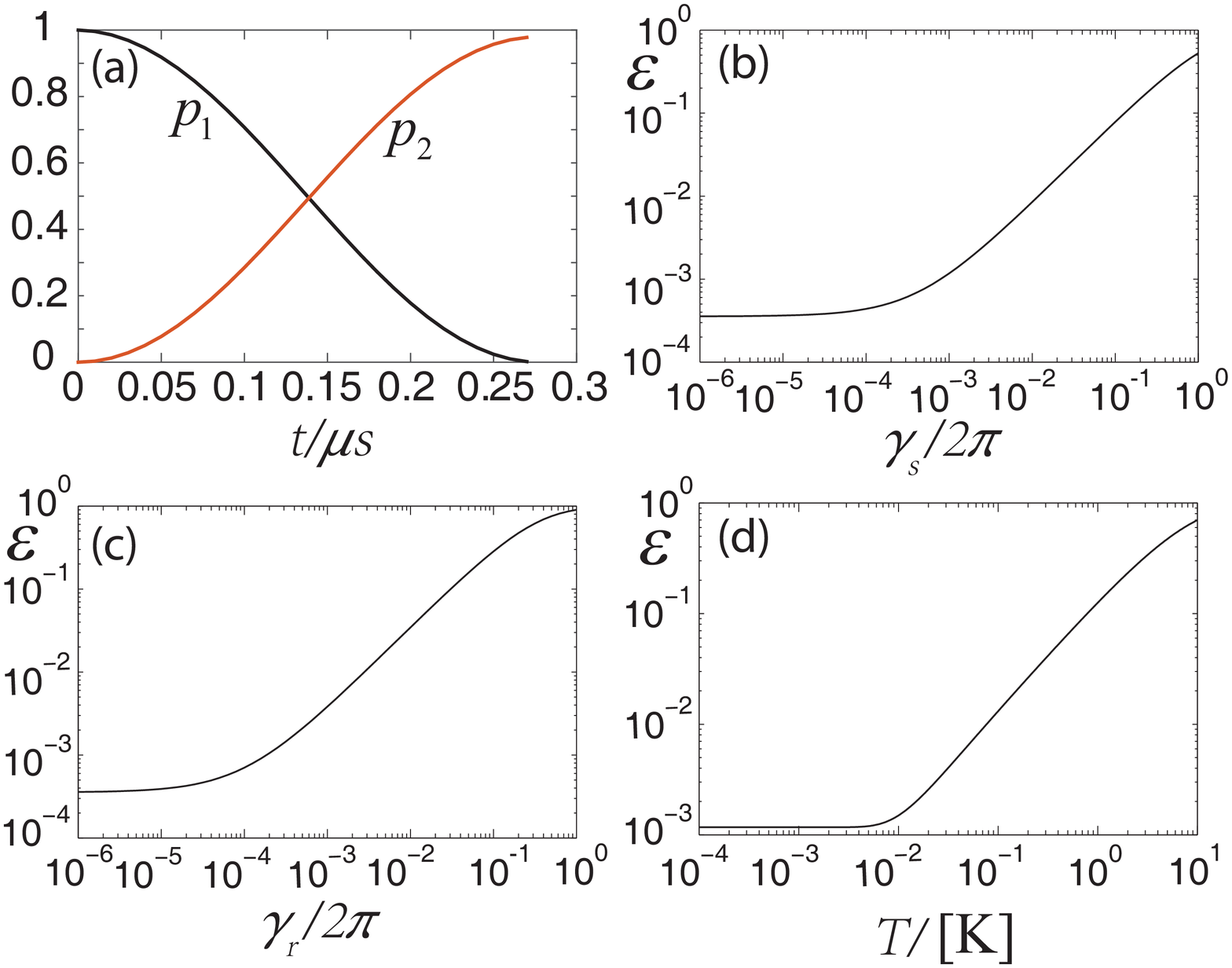}
\caption{\label{fig3}(color online) (a) The time evolution of the elements  of the density operator $p_1(t)=\bra{01}\rho(t)\ket{10}$ and $p_2(t)=\bra{10}\rho(t)\ket{01}$ for initial state $\rho_0=\ket{10}\bra{01}$. The parameters are $(\omega_r,g,\gamma_r,\gamma_s)=2\pi\times(500,0.9, 0.01,0.01)$MHz and  $T=10$mK.  (b) Error $\epsilon$ as a function of the spin dephasing rate $\gamma_s$ for $\gamma_r=0$ and other unchanging parameters. (c)Error as a function of mechanical decoherence rate $\gamma_r$ for $\gamma_s=0$  and  other unchanged parameters. (d) Error as a function of the bath temperature $T$ for $\gamma_s=0$, $\gamma_r=2\pi\times 0.001$MHz, and other unchanging parameters.}
\end{figure}

  \section{Numerical simulations of quantum state transfer}
  To evaluate the system's performance in the anticipated parameter regime, we study the quantum state transfer process of the system with a master equation which includes the unavoidable decoherence. The main source  of decoherence is the  finite lifetime of the qubit spin and the phonon mode as well as the external phonon bath of nonzero temperature $T$. The master equation for the
spin-phonon density operator $\rho(t)$ takes the form
\bea\label{eq7}
\dot{\rho}&=&-\frac{i}{\hbar}[H_{\text{in}},\rho]+(n_B+1)\gamma_r D[a]\rho \notag\\
&+&n_B\gamma_r D[a^\dagger]\rho+\gamma_s D[\sigma_-]\rho,
\eea
where  $n_B=1/(e^{\omega_r/k_BT}-1)$ with the Boltzmann constant $k_B$,  and $D[o]\rho=(o\rho o^\dagger-\frac{1}{2}o^\dagger o\rho-\frac{1}{2}\rho o^\dagger o )$.

We analyzed the fidelities of the quantum state transfer operation by numerically solving Eq.\eqref{eq7}  for initial state $\rho_0=\ket{10}\bra{01}$, for simplicity, where $\ket{mn}$ denotes the spin and phonon  states $\ket{m}$ and $\ket{n}$, respectively. Fig.\ref{fig3}a shows the evolution of the elements  of the density operator $p_1(t)=\bra{01}\rho(t)\ket{10}$ and $p_2(t)=\bra{10}\rho(t)\ket{01}$ when the spin and mechanical qubits are interacting with coupling $g=2\pi\times0.9$ MHz of duration time $t=\pi/2g$.    To find an estimate for the fidelity of the state transfer operation, we calculate the fidelity $F(\rho_t,\rho)=\text {Tr}(\sqrt{\sqrt{\rho}\rho_t\sqrt{\rho}})$ of the desired target state $\rho_t=\ket{01}\bra{10}$ with the actual state $\rho$ that results from the full dynamical evolution. For the experimentally reachable parameters  $(\omega_r,g,\gamma_r,\gamma_s)=2\pi\times(500,0.9, 0.01,0.01)$MHz, and  $T=10$mK we have $F(\rho_t,\rho)=0.979$. Figure \ref{fig3}(b)-(d) shows the resulting state transfer error $\epsilon=1-F$ as a function of  the  mechanical and spin dissipation rates $\gamma_s$ and $\gamma_r$, and the bath temperature $T$, respectively.

\section{CONCLUSIONS}
  We have shown how to electrically control the interaction between an electron spin in a lateral CNT QD and a mechanical motion mode of a CNT. The electrical field displaces the electron density distribution, resulting in the change in the strength of the spin-phonon coupling. This electrically controlled spin-phonon coupling can reach the regime of strong interaction. Numerical simulations show that high-fidelity quantum  state transfer between  spin and phonon qubits is within the reach of the current technology. Potential applications include electrical cooling, state preparation,  and readout of  a nanomechanical resonator \cite{dkdb,acmh,prpc}, nanomechanical-resonator-mediated quantum interface  between optical and spin qubits \cite{kspr,wyrl}, precision measurement \cite{mlob,oapc}, and fundamental tests of quantum theories \cite{ksmr,wmcs}.

 This work was supported by the National Natural Science Foundation of China (11472247 and 61475168), by Zhejiang Provincial Natural Science Foundation of China (Grant No. Y6110314).


\begin{references}
\bibitem{mphz} M. Poot, and H.S.J. van der Zant, Phys. Rep. {\bf 511}, 273 (2012).
\bibitem{bsps} B. Simovi\v{c}, P. Studerus, S. Gustavsson, R. Leturcq, K. Ensslin, R. Schuhmann, J. Forrer, and A. Schweiger, Rev. Sci. Instrum. {\bf77}, 064702 (2006).
\bibitem{fkcb} F.H.L. Koppens,  C. Buizert, K.J. Tielrooij,  I.T. Vink,  K.C. Nowack,  T. Meunier,  L.P. Kouwenhoven, and  L.M.K. Vandersypen, Nature {\bf 442}, 766 (2006).
\bibitem{knfk} K.C. Nowack, F.H.L. Koppens, Yu.V. Nazarov, and L.M.K. Vandersypen, Science {\bf 318}, 1430 (2007).
\bibitem{gsyk} G. Salis, Y. Kato, K. Ensslin, D.C. Driscoll, A.C. Gossard, and D.D. Awschalom,  Nature {\bf414}, 619 (2001).
\bibitem{kfcm}K. Flensberg and C.M. Marcus,  Phys. Rev. B {\bf81}, 195418(2010).
\bibitem{apps}A. P\'{a}lyi, P.R. Struck, M. Rudner, K. Flensberg, and G. Burkard,  Phys. Rev. Lett. {\bf108}, 206811(2012).
\bibitem{cocs}C. Ohm, C. Stampfer, J. Splettstoesser, and M.R. Wegewijs,   Appl. Rev. Lett. {\bf100}, 143103(2012).
\bibitem{jsjg}J. Samm, J. Gramich, A. Baumgartner, M. Weiss, and C. Sch\"{o}nenberger,  J. Appl. Phys. {\bf115}, 174309 (2014).
\bibitem{appss}See Supplemental Material at http://link.aps.org/
supplemental/10.1103/PhysRevLett.108.206811 for more
details..
\bibitem{srmm} S.M. Reimann and M.Manninen, Rev. Mod. Phys. {\bf 74}, 1283(2002).
\bibitem{asmr} A. Secchi and M. Rontani, Phys. Rev. B {\bf 88}, 125403(2013).
\bibitem{jdwa}J.D. Wall,  Phys. Rev. B {\bf76}, 195307(2007).
\bibitem{fksi} F. Kuemmeth, S. Ilani, D.C. Ralph, and P.L. McEuen,  Nature (London) {\bf452}, 448 (2008).
\bibitem{gsah}G.A. Steele, A.K. H\"{u}ttel, B. Witkamp, M. Poot, H. B. Meerwaldt, L.P. Kouwenhoven, and H.S.J. van der Zant, Science {\bf 325}, 1103 (2009).
\bibitem{ahgs} A.K. H\"{u}ttel, G.A. Steele, B. Witkamp, M. Poot, L.P. Kouwenhoven, H.S.J. van der Zant, Nano Lett. {\bf 9}, 2547 (2009).
\bibitem{sakm}S. Amasha, K. MacLean, I. Radu, D.M. Zumbuhl, M.A. Kastner, M.P. Hanson, and A.C. Gossard, 2006, e-print arXiv:cond-mat/0607110.
\bibitem{rhlk}R. Hanson, L.P. Kouwenhoven, J.R. Petta, S. Tarucha, and L.M.K. Vandersypen, Rev. Mod. Phys. {\bf79}, 1217 (2007).
\bibitem{acmh} A.D. O’Connell, M. Hofheinz, M. Ansmann, R.C. Bialczak, M. Lenander, E. Lucero, M. Neeley,
D. Sank, H. Wang, M. Weides, J. Wenner, J.M. Martinis, and A.N. Cleland, Nature {\bf464}, 697(2010).
\bibitem{dkdb} D. Kleckner and D. Bouwmeester1,  Nature (London) {\bf444}, 75 (2006).
\bibitem{prpc}P. Rabl,P. Cappellaro, M.V. Gurudev Dutt, L. Jiang, J.R. Maze, and M.D. Lukin, Phys. Rev. B {\bf79}, 041302(R)(2009).



\bibitem{wyrl}W. Yao, R.-B. Liu, and L.J. Sham, Phys. Rev. Lett. {\bf95}, 030504(2005).

\bibitem{kspr}K. Stannigel,P. Rabl, A.S. S{\o}rensen, P. Zoller, and M.D. Lukin, Phys. Rev. Lett. {\bf105}, 220501(2010).
\bibitem{oapc} O. Arcizet O, P.F. Cohadon, T. Briant, M. Pinard,  and A. Heidmann Nature {\bf444}, 71(2006).
\bibitem{mlob}M.D. LaHaye, O. Buu, B. Camarota, and K.C. Schwab, Science {\bf304}, 74(2004).


\bibitem{ksmr}K.C. Schwab and M.L. Roukes Phys. Today {\bf58}, 36(2005).
\bibitem{wmcs}W. Marshall, C. Simon, R. Penrose,  and D. Bouwmeester Phys. Rev. Lett. {\bf91}, 130401(2003).





\end{references}
\end{document}